\documentclass{PoS}

\def\lsim{\mathrel{\raise.3ex\hbox{$<$\kern-.75em\lower1ex\hbox{$\sim$}}}}
\def\gsim{\mathrel{\raise.3ex\hbox{$>$\kern-.75em\lower1ex\hbox{$\sim$}}}}

\newcommand{\bmp}{\mathbf p}

\title{The $\Upsilon \to X \gamma$ photon spectrum in the soft-energy region}

\ShortTitle{The $\Upsilon \to X \gamma$ photon spectrum in the soft-energy region}

\author{Pedro Ruiz-Femen\'\i a\thanks{I wish to thank the organizers of EFT09 for the excellent arrangements and warm hospitality.
This work is supported in part by the 
DFG Sonder\-forschungsbereich/Transregio~9 
``Computergest\"utzte Theoretische Teilchenphysik'',
by the European Commission MRTN FLAVIAnet under Contract
No. MRTN-CT-2006-035482, and by the
Ministerio de Ciencia e Innovaci\'on under Grant No. FPA2007-60323.
Preprint no. SFB/CPP-09-39.
}\\
        Institut f{\"u}r Theoretische Physik~E, RWTH Aachen University,
        D--52056 Aachen, Germany\\
        E-mail: \email{ruiz@physik.rwth-aachen.de}}

\abstract{The dominant contribution of the next-to-leading order perturbative QCD corrections to the 
$\Upsilon \to X\gamma$ photon 
energy spectrum for photon energies $\omega \sim m_b \alpha_s$ can be obtained trivially from the knowledge of the corresponding ${\cal O}(\alpha)$ correction to the positronium decay spectrum in the same region. The 
latter can be obtained from the NRQED computation of the orthopositronium photon spectrum, or by applying the threshold expansion to the 1-loop QED amplitude, as it is explained in this talk.}

\FullConference{International Workshop on Effective Field Theories: from the pion to the upsilon \\
		February 2-6 2009\\
		Valencia, Spain}

\begin{document}


Apart from being of physical relevance in itself, the Positronium 
system provides a testing ground for the effective field theory (EFT) concepts and techniques devised 
in the study of radiative decay amplitudes, that could eventually be applied to
the description of the more intricate radiative decays of heavy quarkonia ($Q\bar{Q}\to X \gamma$).
An example of the latter is provided in this talk, where we analyze the region of
photon energies $\omega\sim m\alpha$ ({\it soft-energy} region) in the radiative decay of a heavy fermion-antifermion system
through the study of
the $3\gamma$ decay of 
the Positronium spin triplet ground state (orthopositronium: o-Ps). In particular we show that
in this regime
the next-to-leading order (NLO) corrections to the direct decay photon spectrum of a color singlet
$Q\bar{Q}$ pair in angular-momentum configuration $^3 S_1$  are just given by the corresponding
correction to the o-Ps photon spectrum up to a trivial color factor.
The soft-energy 
region is accessible in the o-Ps$\to 3\gamma$ decay because one of the final state photons can 
have an arbitrarily small energy, the other two being hard photons with
energy $\lsim m$. 
The process can be then viewed as the radiative version of the o-Ps$\to X$ decay, where $X=2\gamma$.

The computation of the ${\cal O}(\alpha)$ corrections to the o-Ps spectrum in the soft-energy region can be
considerably simplified using EFT methods or through the asymptotic expansion of the 
relevant 1-loop diagrams. While the latter technique generalizes straightforwardly to the corresponding
computation for quarkonium radiative decays, the EFT framework naturally explains
the origin of the dominant term as a consequence of the non-relativistic nature of the system. 

\section{NRQED computation of the o-Ps decay spectrum for $\omega \ll m$} 
\label{sec:NRQED}

\begin{figure}
\vspace{-0.4cm}
\begin{center}
\includegraphics[width=.25\textwidth]{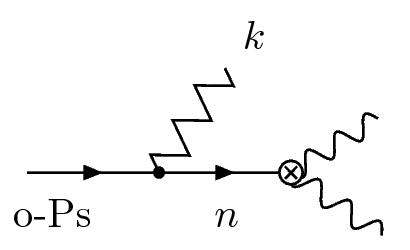}
\end{center}
\label{fig:NRQEDgraph}
\caption{NRQED graph for the o-Ps$\to 3\gamma$ decay. The zig-zag line represents the low-energy photon.}
\end{figure}

The Non-Relativistic QED (NRQED) framework used in Refs.~\cite{Manohar:2003xv,RuizFemenia:2008zz} provides a systematic way to compute 
bound state effects in the the o-Ps decay spectrum when the
photon energy is much smaller than the electron mass.
Contrary to the usual relativistic approach where the 3-photon annihilation is considered to take
place at very short distances as compared to the range of the electromagnetic binding force between $e^+e^-$,
the non-relativistic description takes into account that there is a long-distance part 
in the o-Ps decay process when one the photons in the final state is not
hard ($\omega\ll  m$). In the latter case, the decay proceeds in two steps: 
the low-energy photon is first radiated from the bound state, making a
transition from the C-odd ground state o-Ps (${}^3S_1$) to a C-even Positronium state $n$,
which subsequently decays into two photons (see the EFT graph of Fig.~\ref{fig:NRQEDgraph}). 
The emission of the low energy photon from o-Ps is
described by the Coulomb Hamiltonian of the $e^+e^-$ system 
in interaction with a quantized electromagnetic field~\cite{Manohar:2003xv,RuizFemenia:2008zz}.

Since the computation of the photon spectrum in Ref.~\cite{Manohar:2003xv} was intended
for photon energies comparable to the Positronium binding energy, $\omega\sim m\alpha^2$, 
the interaction Hamiltonian was used in the dipole approximation limit. This approximation
amounts to evaluating the photon vector potential $\mathbf{A}$ in the center of mass of the 
Positronium system, which is fully legitimate for radiated photons with wavelengths
much larger than the characteristic size of the Positronium atom ($a=2/m\alpha$). 
Higher order multipoles arise as a Taylor series in the relative coordinate $\mathbf{x}$ and 
are suppressed by powers of $\omega/a$ under the assumption that the relevant amplitudes 
one has to compute involve integrations to spatial extents of order $\sim a$. 
For photons with larger energies,
$\omega\gg m\alpha^2$, this premise will invalidate the use of the
multipole expansion. 

However, in the case of the o-Ps system that undergoes a radiative transition
before decaying, and as it was properly pointed out by Voloshin~\cite{voloshin}, 
the scale $1/a$ does not constraint the maximum photon energy for which we can
apply the multipole expansion. The reason is that after the soft photon is radiated we have to
consider all possible $e^+e^-$ intermediate states with the right quantum numbers. The 
propagation of these $e^+e^-$ states is described by the Green function obtained from 
the Coulomb Hamiltonian,
\begin{eqnarray}
\left( H_C + {\kappa^2 \over m} \right)   G\left( \mathbf{x}, \mathbf{y}, \kappa \right) = \delta\left( \mathbf{x} - \mathbf{y} \right)\,,
\end{eqnarray}
at energy $\kappa^2/m \simeq  \omega-E_0$.
Since the annihilation of the 
intermediate $e^+e^-$ pair into two photons takes place at small distances,
the amplitude for the full process is thus given by a convolution of the o-Ps
ground state and the Green function. In coordinate space, the Green function
has a characteristic size ruled by the exponential factor $\exp(-\kappa r)$. 
We note that for soft photons of energy $m\alpha^2\ll \omega \ll m$, this exponential 
factor constrains the relevant integration region to distances of order 
$\kappa^{-1}\sim  1/\sqrt{m \omega}$, much smaller than the spatial extent
of the initial Positronium atom. Therefore, the characteristic distance that
enters in the multipole expansion of the soft-photon emission in
the o-Ps$\to 3\gamma$ decay amplitude is determined by the 
falloff of the intermediate Green function rather than by the size of
the Ps atom, so the series of multipoles is actually 
an expansion in powers of $\omega r\sim \sqrt{\omega/m}$ which 
can be used as long as $\omega\ll m$. Note also that for the intermediate
pair, an iterative computation of the Coulomb Green function,
$G=G^{f}+G^{f} V_C G^{f}+\dots$,
with $G^{f}$ the free Green function,
shows that adding a Coulomb exchange generates a term 
$$\int d^3\mathbf{x} \,V_C\,G^{f} \sim {\alpha m\over \kappa}\sim \alpha\sqrt{{m \over \omega}}\ll 1\,,
\quad \mbox{for}\quad m\alpha^2\ll \omega\ll m\,,
$$
as $G^{f}\sim m\exp(-\kappa r)/r$.
Therefore the Coulomb interaction can be treated perturbatively in the 
virtual $e^+e^-$ system after soft photon radiation.  The soft-energy region
$m\alpha^2\ll \omega \ll m$ 
thus separates the regime where binding effects become essential,
from the hard-energy region, where the details of the bound state dynamics are irrelevant.

The NRQED result for the photon energy spectrum in this region, up to NLO reads~\cite{RuizFemenia:2008zz,voloshin}:
\begin{eqnarray}
{ {\rm d}\Gamma_{\rm oPs} \over  {\rm d}x }  &=&{2 m \alpha^6 \over 27 \pi} 
\left[ 5x -\frac{14}{3}\,\alpha\sqrt{x} +  {\cal{O}}(\alpha x) \right]
\,,
\label{eq:1}
\end{eqnarray}
with $x=\omega/m$. The LO term of the NRQED spectrum matches the $\omega \to 0$ limit of the tree-level QED calculation first done by Ore and Powell~\cite{orepowell}. The $\alpha\sqrt{x}$ correction also agrees with  the leading term in the 
$\omega\to 0$ limit of the 1-loop QED result, as shown in Ref.~\cite{RuizFemenia:2008zz} by expanding the analytic expression of the exact 1-loop phase-space distribution given by Adkins~\cite{adkins}.
%

\section{Threshold expansion of the o-Ps$\to 3\gamma$ 1-loop diagrams}
\label{sec:threshold}

\begin{figure}[!t] %
\begin{center}
\vspace{-0.4cm}
\includegraphics[width=.6\textwidth]{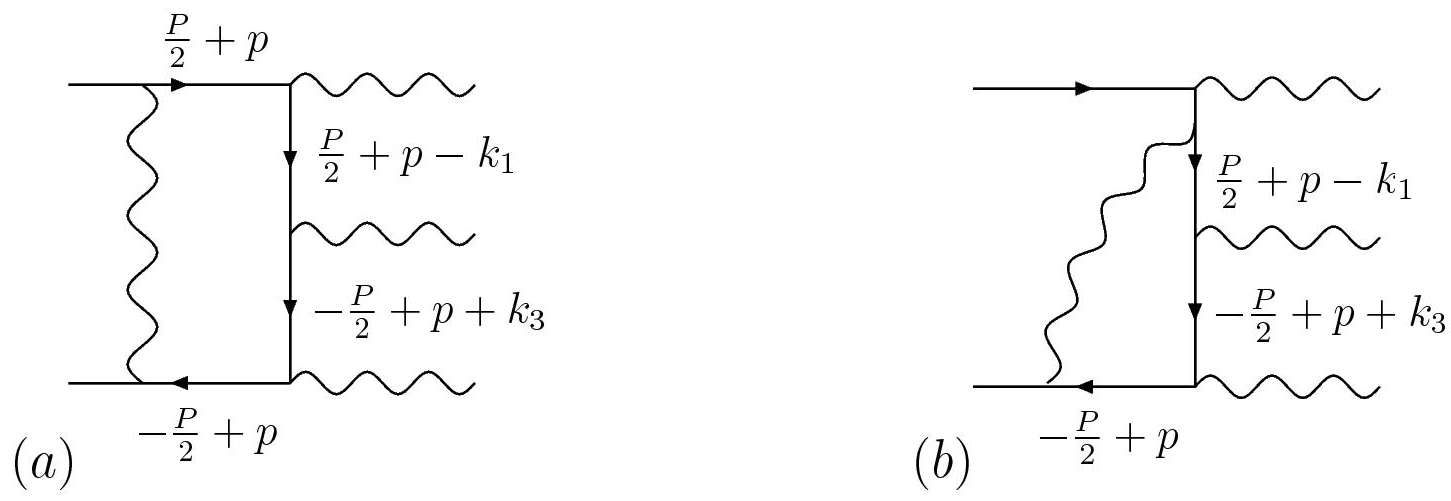}
\label{fig:ladder}
\caption{Ladder and double-vertex graphs contributing to the o-Ps decay amplitude
to order $\alpha$.}
\end{center}
\end{figure}
The NLO term in the NRQED result above arises from
distances in the $e^+e^-$ system of order $r~\sim 1/\sqrt{m\omega}$, that are
much smaller than the size of the Ps atom $r\sim 1/m\alpha$. Translated into momentum
space, this implies that there is a kinematic region where the relative 3-momentum
of the non-relativistic $e^+e^-$ pair obeys the non-relativistic relation $p^0\sim\bmp^2/m\sim\omega$. 
This scaling does not correspond to any of the known modes that have 
been identified in the common applications of non-relativistic EFT's, and
should be compared to the scaling of 
the familiar potential modes, $p^0\sim \bmp^2/m \sim m\alpha^2$, characteristic of the heavy particles that form the
bound state, and that of the soft modes, $p^0\sim |\bmp|\sim m\alpha\gg\bmp^2/m$.
The new loop-momentum region has to be 
taken into account for a successful application of the
threshold expansion method~\cite{beneke} to QED and QCD
loop diagrams
involving a particle-antiparticle system decaying through 
the emission of a soft photon. In the conventional loop expansion,
soft photon radiation from a heavy particle-antiparticle system introduces
a soft energy component $\omega\sim m\alpha$ in the zero component of the heavy particle momenta.
Non-relativistic poles in
massive propagators of the form $(p^0-\bmp^2/2m-\omega)^{-1}$ can be thus found in
these loops, giving rise to a contribution from the loop-momentum region 
$p_0\sim \bmp^2/m \sim \omega$, which we refer to as {\it soft-radiation} region.

To illustrate the latter consider the threshold expansion to the 1-loop o-Ps$\to 3\gamma$ amplitude when 
one of the photons has an energy $\omega_1\sim m\alpha$, where the conventional QED perturbative expansion is still valid.
For our purposes we can take the electron and positron inside the bound state to be at rest. Then $\omega_1$ represents
the smallest scale in the diagrams, and the leading term in the $\omega_1/m$ 
expansion is given by the ladder and double-vertex graphs shown in Fig.~\ref{fig:ladder}. The ladder amplitude involves the computation of the 5-point scalar integral:
\begin{eqnarray}
I_0  =  \int \!\!
{ [d^Dp] \over  p^2 \, [(p+P/2)^2-m^2] \, [(p-P/2)^2-m^2] \,  [(p+P/2-k_1)^2-m^2] \, [(p-P/2+k_3)^2-m^2] }\,,
\nonumber
\end{eqnarray}
with $P=(2m,0)$ the Positronium momentum.
When the loop momentum is hard ($p\sim m$) the corresponding contribution scales as $I_0^{(\rm{h})}\sim m^{-6}$. However, for $\omega_1\to 0$ the loop-momentum regions where $p\ll m$ give larger contributions. Expanding the propagators for $p\ll m$ we have
\begin{eqnarray}
I_0^{(\rm{small})}  = -\frac{1}{2m^2} \int 
{ [d^Dp] \over  (p_0^2-\mathbf{p}^2) \, (2m p_0-\mathbf{p}^2) \, (-2m p_0-\mathbf{p}^2) \,
(2m p_0-2m\omega_1-\mathbf{p}^2) \ }\,,
\nonumber
\end{eqnarray}
which receives contributions from the poles at $p_0=(-\bmp^2/2m+i\epsilon)$ and $p_0=(-|\bmp|+i\epsilon)$ when we close the integration contour in the upper complex $p_0$-plane. From the residue of the former pole we obtain the contribution from the soft-radiation region 
where $p_0\sim\mathbf{p}^2/m\sim m\omega_1$:
\begin{eqnarray}
I_0^{(\rm{s-r})}  = -\frac{i}{16 m^3} \int [d^n\bmp] \,
{1 \over  (\mathbf{p}^2)^2 \, (\mathbf{p}^2+m\omega_1)  \,  } = \frac{1}{64\pi m^3} (m\omega_1)^{-\frac{3}{2}}\,,
\nonumber
\end{eqnarray}
which gives the leading term in the $\omega_1/m$ expansion of $I_0$. The pole at $p_0=(-|\bmp|+i\epsilon)$ gives an integral dominated by $\bmp\sim m\omega_1$ which scales as 
$[d^Dp]/(m p_0)^5\sim m^{-5}\omega_1^{-1}$, {\it i.e.} larger
than the hard region one, although suppressed by $\sqrt{\omega_1/m}$ with respect the term from the soft-radiation region. 
It is straightforward to verify that only the 1-loop graphs in Fig.~\ref{fig:ladder} receive contributions from the soft-radiation region~(see Ref.~\cite{RuizFemenia:2008zz} for a more detailed discussion). Once propagator numerators and vertex factors are included, the soft-radiation piece in the 
threshold expansion of the 1-loop $e^+e^-\to 3\gamma$ amplitude reproduces the NLO term in the photon spectrum shown in Eq.~(\ref{eq:1}).

\section{The direct photon spectrum in the radiative $\Upsilon$ decay for $\omega \sim m_b \alpha_s$}
\label{sec:quarkonium}

\begin{figure}[!t] %
\begin{center}
\vspace{-0.4cm}
\includegraphics[width=.55\textwidth]{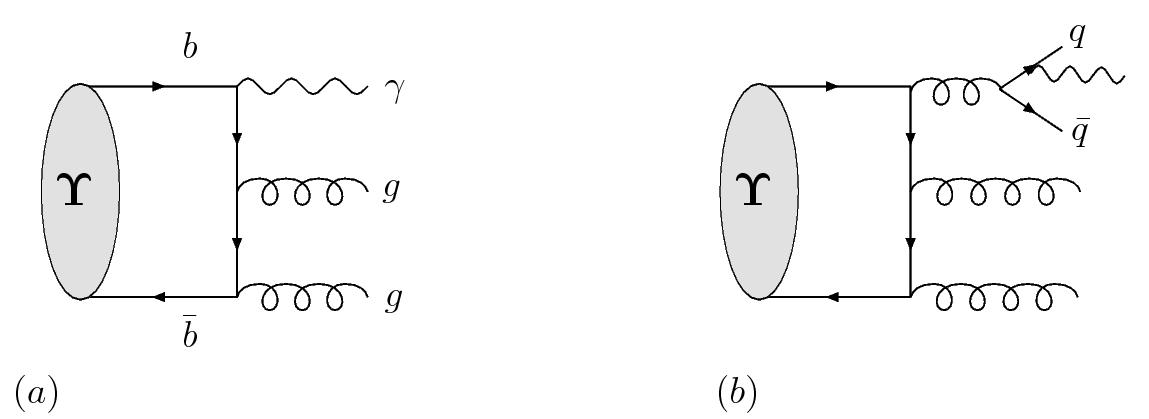}
\end{center}
\vspace{-0.4cm}
\label{fig:Upsilon}
\caption{Leading-order Feynman diagrams for radiative $\Upsilon$ decay.}
\end{figure}
The photon energy spectrum in heavy quarkonium radiative decays receives contributions from two 
different mechanisms. The direct photon spectrum entails those contributions where the observed photon is radiated off a heavy quark, Fig.~\ref{fig:Upsilon}(a), while the fragmentation contributions are those in which one of the partons 
fragments and transfers a fraction of its momentum to the photon,  Fig.~\ref{fig:Upsilon}(b).
The NRQCD factorization approach~\cite{bbl} allows to write down an operator product expansion
for the direct photon spectrum in, for example, $\Upsilon$ decays:
\begin{eqnarray}
{{\rm d}\Gamma_{\rm{dir}} \over {\rm d}x} = \sum_n C_n(x) \langle \Upsilon|{\cal O}_n | \Upsilon\rangle\,,
\end{eqnarray}
where the sum above extends over all $Q\bar{Q}[n]$ configurations that can be found inside
the quarkonium, and the $C_n(x)$
are short distance Wilson coefficients that can be determined from the annihilation
cross section of the on-shell $Q\bar{Q}[n]$ pair as a perturbative series in $\alpha_s(m_b)$.
The ${\cal O}_n$ in the long-distance matrix element are NRQCD operators which are
organized in powers of the relative velocity of the heavy quarks.
At leading order
only the color-singlet operator ${\cal O}_1({}^3S_1)$, that creates and annihilates
a quark-antiquark pair in a color-singlet ${}^3S_1$ configuration, contributes. The
nonperturbative NRQCD matrix element is related to the $\Upsilon$ wave function at the origin
\begin{eqnarray}
\langle \Upsilon|{\cal O}_1({}^3S_1) | \Upsilon\rangle
= \langle \Upsilon | \phi_{\bmp}^\dagger \sigma^i \chi_{-\bmp} \, \chi^{\dagger}_{-\bmp^\prime} \sigma^i \phi_{\bmp^\prime} 
 | \Upsilon\rangle=2N_c\,|\psi(0)|^2
\,.
\nonumber
\end{eqnarray}
The matching coefficient at leading order in $\alpha_s$ comes from the tree-level $Q\bar{Q}\to gg \gamma$
annihilation~\cite{upsilon} ($x=\omega/m_b$ in this case),
\begin{eqnarray}
\!\!\!\!\!\!C_1^{\,\mbox{\scriptsize LO}}(x) &=& {32\over 27}{ \alpha\alpha_s^2 Q^2_b \over m_b^2 } 
\left[\frac{2 - x}{x} 
+ \frac{\left( 1 - x\right) \,x}{{\left( 2 - x \right) }^2} 
- \frac{2\,{\left( 1 - x \right) }^2\,\log (1 - x)}{{\left( 2 - x \right)}^3}
 + \frac{2\,\left( 1 - x\right) \,\log (1 - x)}{x^2} \right]\,,
\end{eqnarray}
which has identical $x$-dependence to that of the Ore-Powell spectrum for o-Ps~\cite{orepowell}.
Indeed the factorization approach applies trivially to QED bound states, where everything
is calculable perturbatively and the bound state wavefunction reduces to the Coulomb one.
\begin{figure}[!t] %
\begin{center}
\includegraphics[width=.8\textwidth]{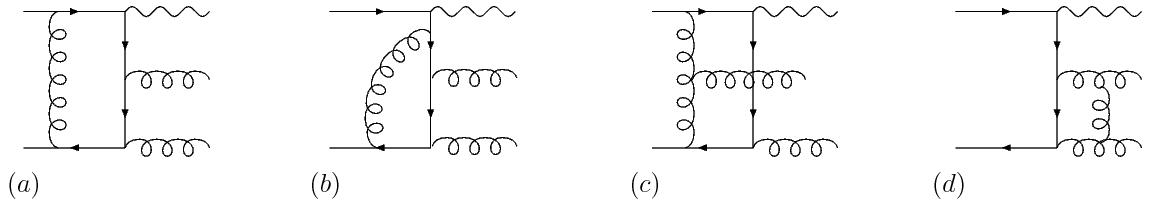}
\end{center}
\vspace{-0.4cm}
\label{fig:Upsilondiagrams}
\caption{A sample of the next-to-leading-order diagrams for the direct contributions to radiative $\Upsilon$ decay.}
\end{figure}

\begin{figure}[t] %
\begin{center}
\hspace{-1cm}
\includegraphics[width=.6\textwidth]{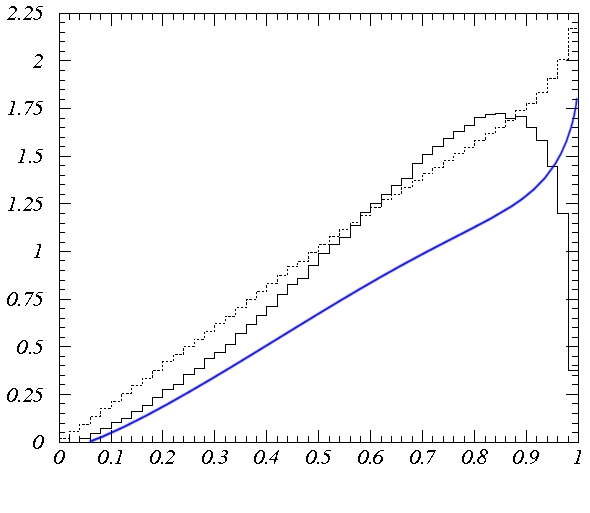}
\put(-300,190){\Large ${1 \over \Gamma_{\rm{dir}}^{\rm{ LO}}} { {\rm d}\Gamma_{\rm{dir}} \over {\rm d}x}$}
\put(-30,5){\large $x$}
\end{center}
\label{fig:spectrum}
\caption{The direct photon energy spectrum in $\Upsilon$ radiative decay. The stepped dotted line is the leading-order direct photon spectrum, and the solid stepped line includes the full ${\cal O}(\alpha_s)$ corrections
from Ref.~\cite{kraemer}. The solid blue line is the 
approximated result given by the leading term in the $\omega/m_b$ expansion of the ${\cal O}(\alpha_s)$ corrections.}
\end{figure}

The threshold expansion technique used above for the o-Ps$\to 3 \gamma$ decay amplitude, can
be applied now to calculate the ${\cal O}(\alpha_s)$ correction to the 
short-distance coefficient $C_1(x)$
in the region where $\omega \sim m_b \alpha_s$. A reduced set of the 1-loop diagrams that
have to be considered for the virtual corrections ($b\bar{b}\to \gamma gg$) is shown in Fig.~\ref{fig:Upsilondiagrams}.
The abelian topologies are obviously the same as those encountered in the $e^+e^-\to 3\gamma$ 1-loop graphs. 
Therefore we know that the leading contribution in the $\omega/m_b$ expansion comes from the soft-radiation region
in the ladder and double-vertex topologies, Figs.~\ref{fig:Upsilondiagrams}(a) and \ref{fig:Upsilondiagrams}(b). 
A quick examination of the pole structure from the propagators in non-abelian diagrams
reveals that they do not receive contributions from the soft-radiation region, and thus give subleading
terms in $\omega/m_b$. Likewise, real corrections corresponding to tree-level diagrams with one more parton in the final state,
$b\bar{b}\to \gamma ggg,\gamma gq\bar{q}$ are not enhanced for $\omega\to 0$. Therefore we can extract the leading ${\cal O}(\alpha_s)$ correction to  $C_1(x)$
when $\omega \sim m_b \alpha_s$  from the 
QED result~(\ref{eq:1}) with trivial modifications: 
\begin{eqnarray}
C_1^{\,\mbox{\scriptsize NLO}}(x) &=& 
 \left (8\over 27\right)^2 { 3\alpha\alpha_s^2 Q_b^2 \over m_b^2 }\left[
-{14\over 3}\,\alpha_s\sqrt{x}  +{\cal O}(\alpha_s x) \right]
\,.
\label{eq:2}
\end{eqnarray}
From the result~(\ref{eq:2}) we obtain a simple expression for the NLO correction to the direct
spectrum in  the soft-energy 
region for the radiative $\Upsilon$ decay:
\begin{eqnarray}
{1 \over \Gamma_{\rm{dir}}^{\,\mbox{\scriptsize LO}}} { {\rm d}\Gamma_{\rm{dir}}^{\,\mbox{\scriptsize NLO}} \over {\rm d}x} 
\,=\,
 -{56 \over 27} {1\over (\pi^2-9)} \alpha_s \sqrt{x}  
\,=\,
\left(4 \over 3\right)\,{1 \over \Gamma_{\rm oPs}^{\,\mbox{\scriptsize LO}}} { {\rm d}\Gamma_{\rm oPs}^{\,\mbox{\scriptsize NLO}} \over {\rm d}x} \,.
\label{eq:3}
\end{eqnarray}
The factor $4/3$ above is just given by the ratio between the colour factors in the ladder  and
tree-level diagrams. 
The full perturbative ${\cal O}(\alpha_s)$ corrections to the $\Upsilon$ photon energy spectrum
were calculated numerically in Ref.~\cite{kraemer}. We have superimposed the NLO result for the spectrum
including the correction~(\ref{eq:3})
to the complete ${\cal O}(\alpha_s)$ result from Ref.~\cite{kraemer} in Fig.~\ref{fig:spectrum}, where $\alpha_s=0.2$. 
As expected, our result can only improve the LO prediction for $x\lsim 0.3$, as
it was obtained assuming the scaling $x\sim \alpha_s$. 
The agreement with the exact result for small $x$ is remarkably good considering that our
formula neglects terms of order $\sqrt{x}$. The approximated NLO direct spectrum becomes negative
for $x\lsim 0.05$. However, at those energies ($x\lsim \alpha_s^2$) the emission of the photon can produce transitions to virtual bound states and does not longer belong to the short-distance part of the decay~\cite{Manohar:2003xv}. 
Moreover, we should be aware that the fragmentation contributions to the photon spectrum
become important in the low-$x$ region~\cite{catani,maltoni} (namely for $x\lsim 0.3$ in $\Upsilon$ decays),
and have to be properly taken into account for a computation of the full spectrum.
The standard NRQCD factorization also 
breaks down at large values of the photon energy~\cite{rothstein}, where one needs to consider also collinear
degrees of freedom~\cite{upsiloncoll}.

\end{document}